%% file: top2016_proceeding.tex
\documentclass[12pt]{article}
\usepackage{graphicx}
\usepackage{subfig}

\newcommand\pubnumber{SNSN-323-63}
\newcommand\pubdate{\today}

\def\institute{}

\def\Title#1{\begin{center} {\Large #1 } \end{center}}
\def\Author#1{\begin{center}{ \sc #1} \end{center}}
\def\Address#1{\begin{center}{ \it #1} \end{center}}

\newcommand\pubblock{\rightline{\begin{tabular}{l} \pubnumber\\
         \pubdate  \end{tabular}}}
\newenvironment{Abstract}{\begin{quotation}  }{\end{quotation}}
\newenvironment{Presented}{\begin{quotation} \begin{center} 
             PRESENTED AT\end{center}\bigskip 
      \begin{center}\begin{large}}{\end{large}\end{center} \end{quotation}}

\input econfmacros.tex

\begin{document}
\begin{titlepage}
\pubblock

\vfill
\Title{Measurement of the W boson helicity using top pair events at $\sqrt{s}$ = 8 TeV with the CMS detector}
\vfill
\Author{ Mohsen Naseri 
\\
on behalf of the CMS Collaboration}
\Address{\institute School of Particles and Accelerators, Institute for Research in Fundamental\\
Sciences(IPM),\\
P. O. Box 19 56 83 66 81, Tehran, Iran
}
\vfill
\begin{Abstract}
This document gives an overview over the recent results on helicity measurement of W boson originated from top pair events. The results are obtained using data collected by the CMS detector at a center-of-mass energy of 8 TeV. The helicity measurements are confronted with the most precise theoretical predictions of the standard model.
\end{Abstract}

\vfill
\begin{Presented}
$9^{th}$ International Workshop on Top Quark Physics\\
Olomouc, Czech Republic,  September 19--23, 2016
\end{Presented}
\vfill
\end{titlepage}
\def\thefootnote{\fnsymbol{footnote}}
\setcounter{footnote}{0}

\section{Introduction}
Top quarks decay almost exclusively into a b quark and a W boson via the electroweak interaction. In particular, the measurement of the W boson polarization in the top quark decays allows to probe the tWb structure and to search for possible extensions of the standard model (SM).
 
In general, W bosons in the top quark decays can be produced in three states of left-handed, right-handed, and longitudinal helicity. Since the W boson couples to a b quark of left-handed chirality which translates into left-handed helicity in the massless limit of the b quark, right-handed W bosons are not expected to be produced in the top quark decays.
Defining $\Gamma_{L, 0, R}$ as the partial width of the top quark decaying into left-handed, right-handed, and longitudinal W boson helicities, the helicity fractions are given by $F_{L, 0, R}=\frac{\Gamma_{L, 0, R}}{\Gamma_{total}}$.

The W boson polarization affects several kinematic variables in which can be used to measure the helicity components. Among all relevant kinematic observables which are sensitive to the W boson helicity fractions, the widely used one is the angular distributions of the top quark decay products. All following measurements employs this observable to extract the helicity fractions.
\section{Measurement of the W boson helicity using $t\bar{t}$ events in the dilepton final state at $\sqrt{s}$ = 8 TeV}
\begin{figure}[thp]
\begin{center}
     \subfloat[]{\includegraphics[width=0.48\textwidth , height=0.4\textwidth]{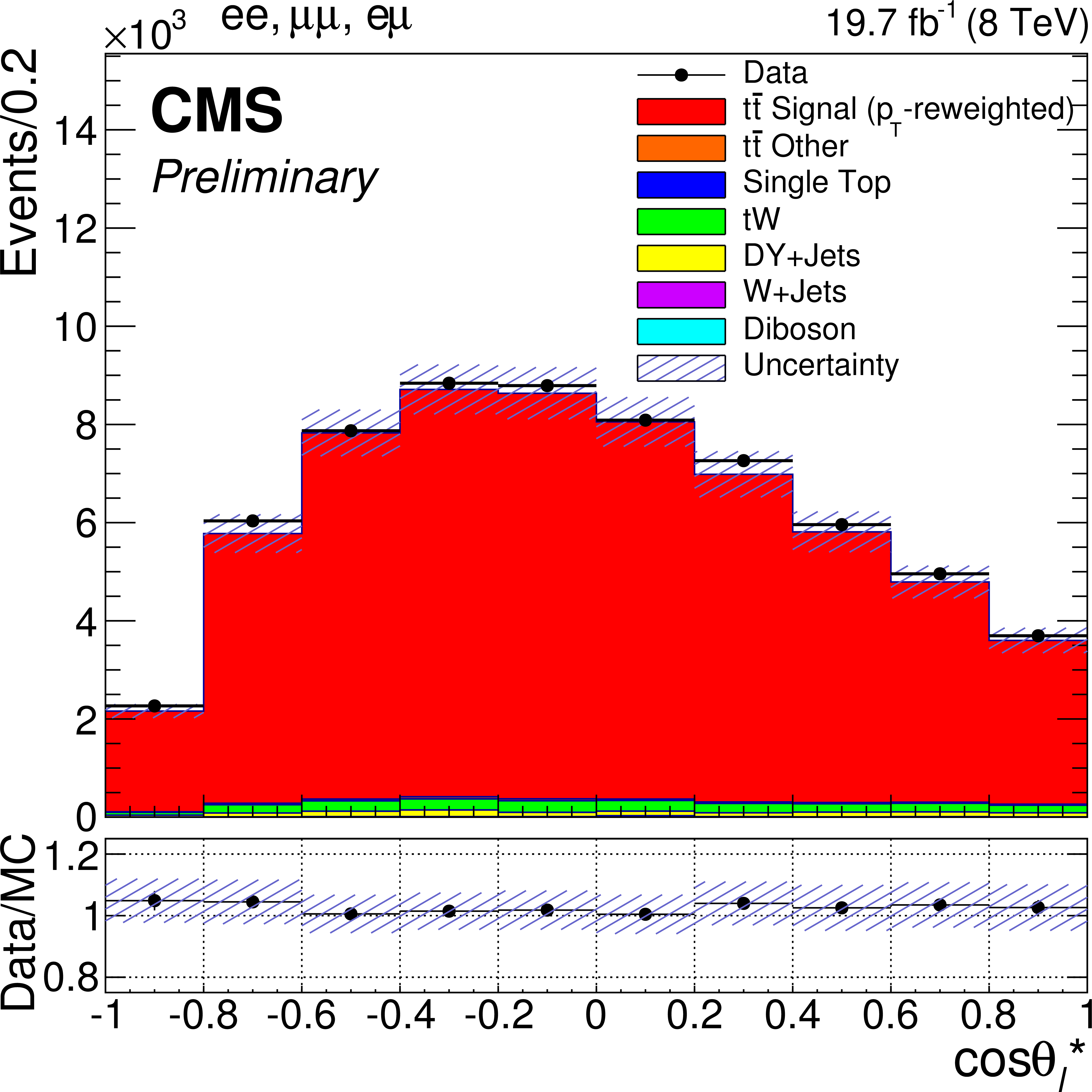}}
     \subfloat[]{\includegraphics[width=0.48\textwidth , height=0.4\textwidth]{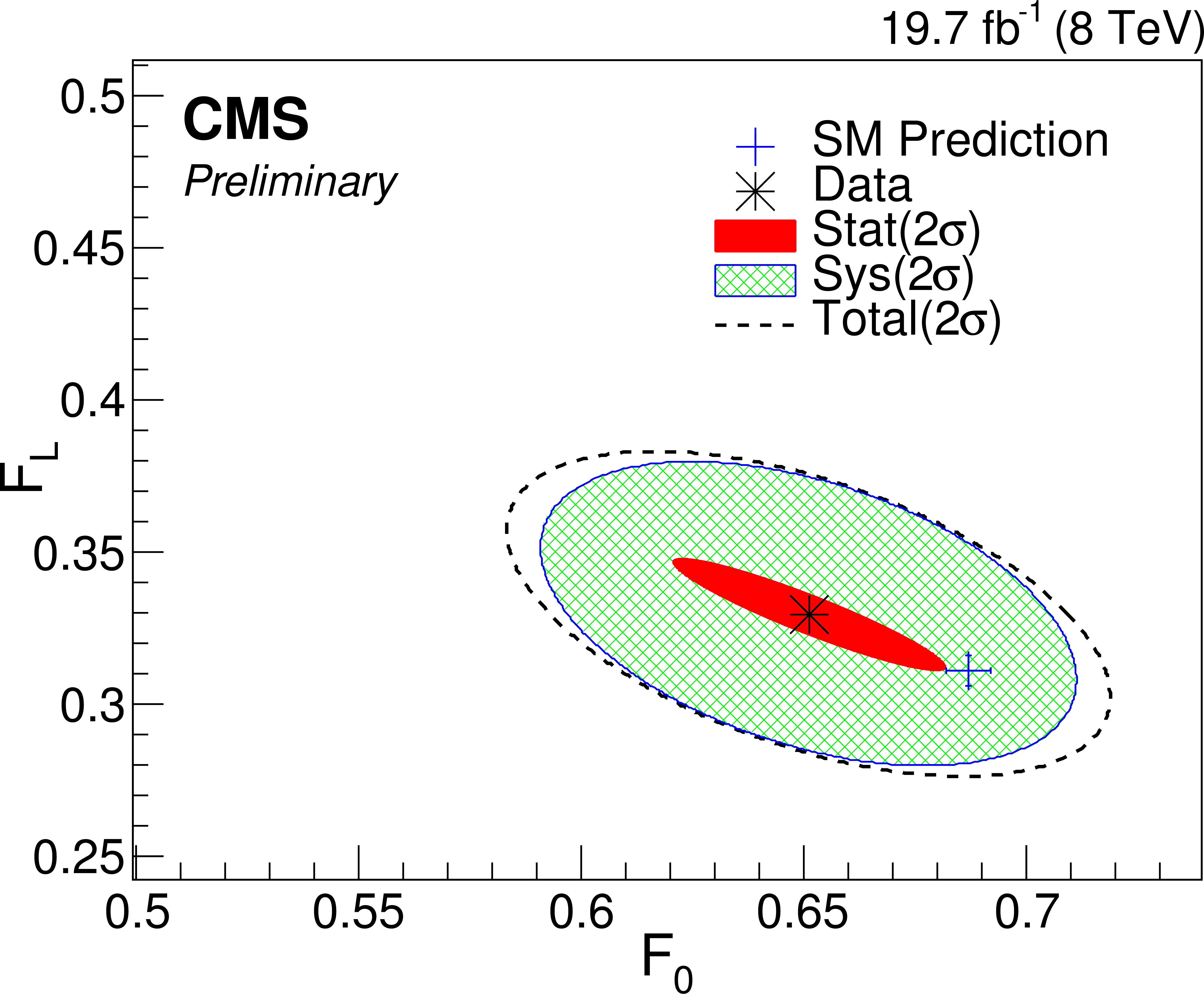}}
     \caption{(a) Distribution of the cos($\theta^{*}$) for the three dilepton channels considered together. (b) The 95$\%$ region in the (F$_{0}$, F$_{L}$) plane obtained from the fit to data. The measured and theoretical values of the W boson helicity fractions are shown as well \cite{CMS-PAS-TOP-14-017}.\label{fig::generalcontour}}
\end{center}
\end{figure}

The first analysis presented uses $t\bar{t}$ events with two leptons, electrons and/or muons, in the final state~\cite{CMS-PAS-TOP-14-017}.
The analysed data sample corresponds to an integrated luminosity of 19.7 fb$^{-1}$ at a center of mass energy of 8 TeV, collected by the CMS detector \cite{Chatrchyan:2008aa}.  
Events are required to contain two charged leptons with opposite sign, missing transverse energy, and two b tagged jets. Background originating from Drell-Yan (DY) events is suppressed by requiring large missing energy in the $e^{+}e^{-}$ and $\mu^{+}\mu^{-}$ channels. In addition, dimuon or dielectron events in the region around the Z boson mass peak are also rejected.
The contribution of DY+jets events in dimuon and dielectron is estimated from a Z boson mass window control region,  and is used to normalize the simulation in the signal region. A analytical Matrix Weighting Technique (AMWT)~\cite{Abbott:1997fv} is used to reconstruct best top pair candidates. The cos($\theta^{*}$) distribution, which is used to perform the measurement, is presented in Figure~\ref{fig::generalcontour}.

In order to extract the W boson helicity fractions, a reweighting technique as explained in \cite{Chatrchyan:2013jna} is used. In this method, the reweighted signal distribution of cos($\theta^{*}$) in simulation is fitted to the observed distribution. The W boson helicity fractions, obtained from a fit to the reconstructed distributions of cos($\theta^{*}$),  are F$_{L}$ = 0.329 $\pm$ 0.029, F$_{0}$ = 0.653 $\pm$ 0.026, and F$_{R}$ = 0.018 $\pm$ 0.027.

\section{Measurement of the W boson helicity using $t\bar{t}$ events in the semi-leptonic final state at $\sqrt{s}$ = 8 TeV}
The CMS collaboration also reports the study of the W-boson helicity fractions in top-quark decays using a sample of $t\bar{t}$ events where one of the top quarks decays semileptonically and the other decays hadronically~\cite{Khachatryan:2016fky}.
The analysis is done using the collected data in 2012 with the CMS detector at the LHC, corresponding to an integrated luminosity of 19.8 fb$^{-1}$. The event selection requires either one muon or one electron, along with four jets in the final state in which two of them must be identified as originating from b quarks. Events with an additional soft muon or and additional soft electron are vetoed in order to reject backgrounds from dileptonic $t\bar{t}$ and Drell–Yan events. To reduce the QCD multijet background, the transverse mass of the leptonically decaying W boson, is required to be greater than 30 GeV/c.

 A kinematic fit is used to determine the best combination of b jets, other jets, and lepton candidates to the top quark and antiquark decay hypotheses. The reconstructed helicity angle distributions are then fitted to measure the W-boson helicity fractions and to derive possible anomalous tWb couplings. 

\begin{figure}[thp]
\begin{center}
     \subfloat[]{\includegraphics[width=0.48\textwidth , height=0.4\textwidth]{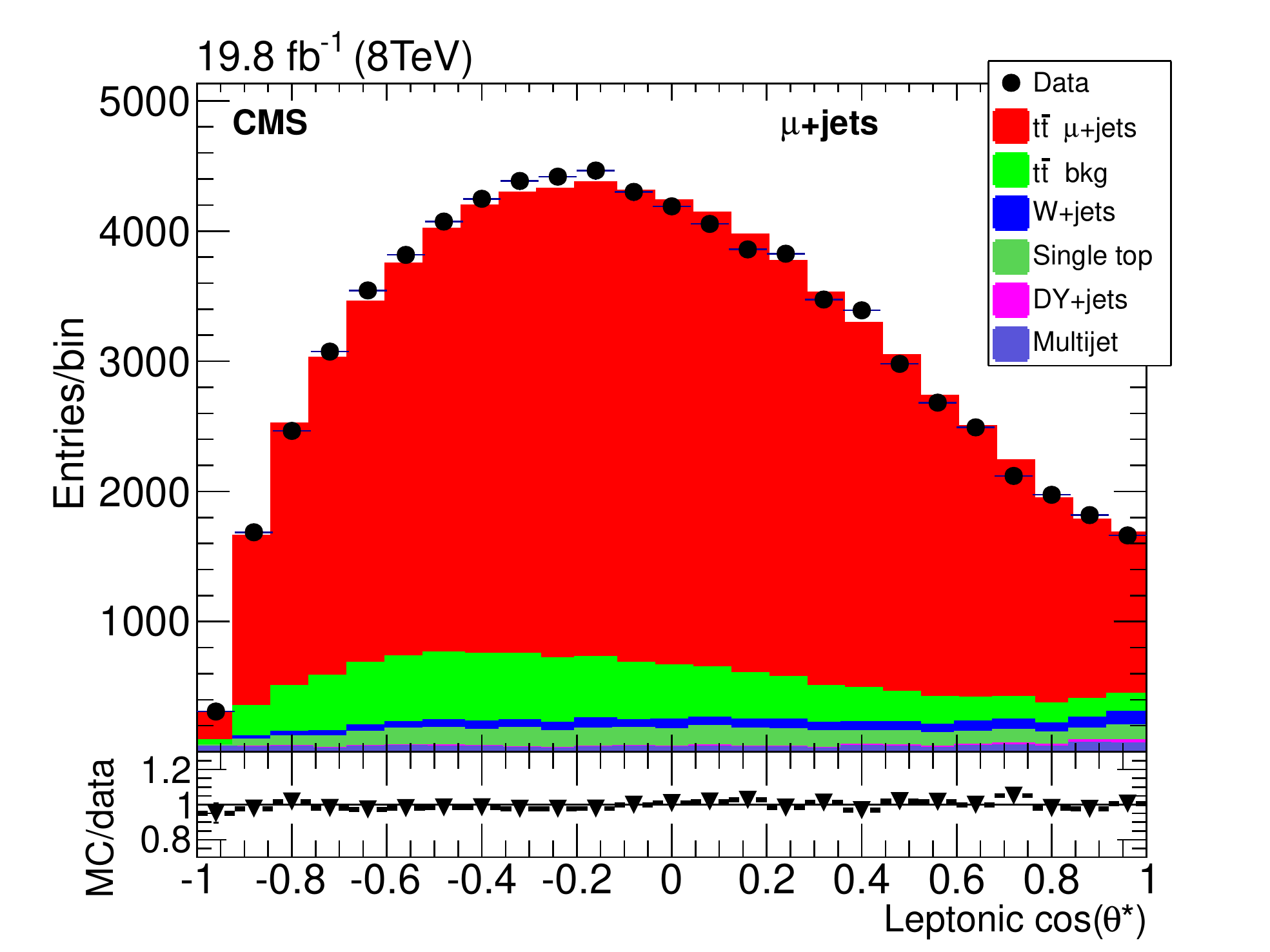}}
     \subfloat[]{\includegraphics[width=0.48\textwidth , height=0.4\textwidth]{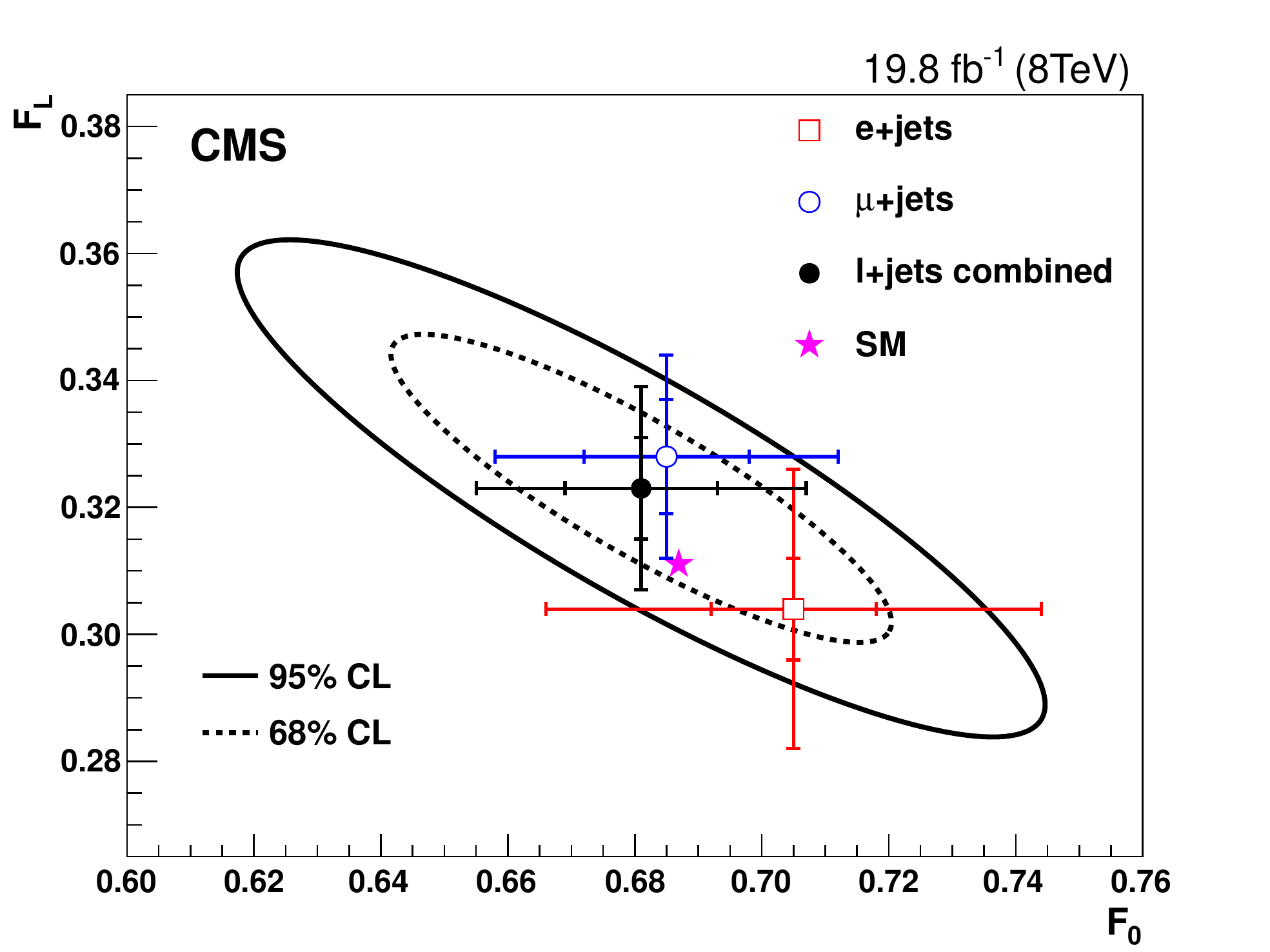}}
     \caption{(a) Distribution of the cos($\theta^{*}$) in the leptonic branch. (b) The measured W boson helicity fractions in the (F$_{0}$, F$_{L}$) plane obtained from the fit to data \cite{Khachatryan:2016fky}.\label{fig::semicontour}}
\end{center}
\end{figure}

Figure \ref{fig::semicontour}(a) shows the distribution for the cos($\theta^{*}$) of the helicity angle from the leptonic $\mu$+jets branch.
The measured W boson helicity fractions are found to be F$_{0}$ = 0.681 $\pm$  0.012 (stat.) $\pm$ 0.023 (syst.), F$_{L}$ = 0.323 $\pm$ 0.008 (stat.) $\pm$ 0.014 (syst.), and F$_R$ = - 0.004 $\pm$ 0.005 (stat.) $\pm$ 0.014 (syst.), which are consistent with the SM expectations. Figure \ref{fig::semicontour}(b) shows the measured W boson helicity fractions in the (F$_{0}$, F$_{L}$) plane with the allowed two-dimensional 68$\%$ and 95$\%$ CL regions.

\end{document}

%% file: econfmacros.tex
\def\beq{\begin{equation}}
\def\eeq#1{\label{#1}\end{equation}}
\def\eeqn{\end{equation}}

\def\beqa{\begin{eqnarray}}
\def\eeqa#1{\label{#1}\end{eqnarray}}
\def\eeqan{\end{eqnarray}}

\let\bar=\overbar

\def\Dslash{\not{\hbox{\kern-4pt $D$}}}
\def\dslash{\not{\hbox{\kern-2pt $\del$}}}

\def\msb{{\bar{\ssstyle M \kern -1pt S}}}

